\documentclass{iopart}             
\usepackage[T1]{fontenc}
\usepackage[applemac]{inputenx}
\usepackage{graphicx}
\usepackage{url,graphics,epsfig}
\usepackage{amssymb}

\usepackage{tikz}

\newcommand*\mycirc[1]{%
  \begin{tikzpicture}[baseline=(C.base)]
    \node[draw,circle,inner sep=1pt](C) {#1};
  \end{tikzpicture}}

\begin{document}

\jl{2}
%
%
%
\def\etal{{\it et al~}}
\def\newblock{\hskip .11em plus .33em minus .07em}
%
%
%
%
%
%
\setlength{\arraycolsep}{2.5pt}             

\title[{K-Shell Photoionization of  B-like Atomic Nitrogen Ions}]{{\it K}-Shell Photoionization
	of  B-like Atomic Nitrogen Ions: Experiment and Theory}

\author{  M F Gharaibeh$^{1}$, N El Hassan$^{2}\footnote[2]{Present address: Laboratoire Structures, Propri\'{e}t\'{e}s et Mod\'{e}lization 
              des Solides (SPMS) UMR CNRS 8580, Ecole Centrale Paris, 1, Grande Voie des Vignes, 92295 Ch\^{a}tenay-Malabry, France}$,  
              M M Al Shorman$^{2}$, J M Bizau$^{2,3}\footnote[1]{Corresponding author, E-mail: jean-marc.bizau@u-psud.fr}$,  
              D Cubaynes$^{2,3}$, S Guilbaud$^{2}$,  I Sakho$^{4}$, C Blancard$^{5}$ 
              and B M McLaughlin$^{6,7}\footnote[3]{Corresponding author, E-mail:b.mclaughlin@qub.ac.uk}$}

\address{$^{1}$Department of Physics, Jordan University of Science and Technology, Irbid 22110, Jordan}

\address{$^{2}$Institut des Sciences Mol\'{e}culaires d'Orsay (ISMO), CNRS UMR 8214, 
			Universit\'{e} Paris-Sud, B\^{a}t. 350, F-91405 Orsay cedex, France}

\address{$^{3}$Synchrotron SOLEIL - L'Orme des Merisiers, Saint-Aubin - BP 48 91192 Gif-sur-Yvette cedex, France}

\address{$^{4}$Department of Physics, UFR of Sciences and Technologies, 
                           University Assane Seck of Ziguinchor, Ziguinchor, Senegal}

\address{$^{5}$CEA-DAM-DIF, Bruy$\grave{\rm e}$res-le-Ch\^{a}tel, F-91297 Arpajon Cedex, France}

\address{$^{6}$Centre for Theoretical Atomic, Molecular and Optical Physics (CTAMOP),\\
			School of Mathematics and Physics, The David Bates Building, 7 College Park, Queen's University Belfast, 
			Belfast BT7 1NN,  UK}

\address{$^{7}$Institute for Theoretical Atomic and Molecular Physics (ITAMP),\\
			Harvard Smithsonian Center for Astrophysics, MS-14, Cambridge, MA 02138, USA}


%
%

\begin{abstract}
Measurements of absolute cross sections for the {\it K}-shell photoionization of B-like atomic nitrogen ions 
were carried out utilizing the ion-photon merged-beam technique at the SOLEIL synchrotron radiation facility in 
Saint-Aubin, France.  High-resolution spectroscopy with E/$\Delta$E $\approx$ 13,500
 was the maximum resolution achieved. We have investigated two photon energy regions: 
 404 eV -- 409 eV and 439 eV -- 442 eV.  Resonance peaks 
 found in the experimental measured cross sections are compared with theoretical estimates from the 
 multi-configuration Dirac - Fock, R-matrix and empirical methods, allowing 
 identification of the strong 1s  $\rightarrow$ 2p and the weaker 
1s $\rightarrow$ 3p resonances in the observed {\it K}-shell spectra of this B-like nitrogen ion.  
\end{abstract}

%
%


\vspace{1.0cm}
\begin{flushleft}
Short title: {\it K}-shell photoionization of B-like atomic nitrogen ions\\
\submitto{\jpb:~\today}
\end{flushleft}

\maketitle
%

%
%
\section{Introduction}
X-ray spectra from {\it XMM-Newton} may be utilized to characterize the interstellar medium (ISM), 
if accurate atomic {\it K}-edge cross sections are available  
\cite{McLaughlin2001, Brickhouse2010, Kallman2010, Quinet2011,Soleil2011,Soleil2013,McLaughlin2013}. 
Single and multiply ionization stages of C, N, O, Ne and Fe have been observed 
in the ionized outflow in  NGC 4051 measured with {\it XMM-Newton} \cite{Olge2004} 
in the soft x-ray region and low ionized stages of C, N and O have also been used 
in modelling x-ray emission from OB super-giants \cite{Cassinelli1981}.
Radiative/photo recombination of singly and doubly charged nitrogen ions also 
play an important role in the chemistry of the atmosphere of Titan \cite{Dutuit2013}.
Detailed photoionization models of the brightest knot of star formation in the blue
compact dwarf galaxy Mrk 209 required abundances for ions of oxygen and nitrogen \cite{Diaz2007}.
The  XMM-{\it Newton} X-ray spectra of WR 1 is rich in nitrogen ions  \cite{Ignace2003} and
photoionization cross section data and abundances for carbon, nitrogen, and oxygen in their various stages  of ionization are 
essential for photoionization models applied to the plasma modelling in a variety of planetary nebulae \cite{Bohigas2008}. 
In the present study we focus our  attention on obtaining detail spectra on the 
doubly ionized nitrogen ion N$^{2+}$ (N III) in the vicinity of its {\it K} - edge.  

Photoionization (PI) cross sections used for  the modelling of astrophysical
phenomena have mainly been provided by theoretical methods,
due to limited experimental data being available. Major effort has gone into improving the
quality of calculated data using state-of-the-art theoretical methods. 
Recent advances in the determination of atomic parameters for modeling {\it K} lines
in cosmically abundant elements have been reviewed by Quinet and co-workers \cite{Quinet2011}.

Absolute experimental {\it K}-shell photoionization cross section results have been 
obtained by various groups on a variety of atoms and ions of astrophysical interest;
He-like Li$^{+}$ \cite{Scully2006,Scully2007,DPI2013},
Li-like  B$^{2+}$ \cite{Mueller2010}, C$^{3+}$  \cite{Mueller2009}, N$^{4+}$ \cite{Soleil2013},
Be-like B$^{+}$ \cite{Mueller2014}, C$^{2+}$ \cite{Scully2005}, N$^{3+}$ \cite{Soleil2013},
B-like C$^{+}$ \cite{Schlachter2004},
C-like N$^{+}$ \cite{Soleil2011},
N-like O$^{+}$ \cite{Kawatsura2002},
F-like Ne$^{+}$ \cite{Yamaoka2001},
neutral nitrogen \cite{McLaughlin2011} and oxygen \cite{Krause1994,Menzel1996,Stolte1997,Stolte2013}.

Recent studies on {\it K}-shell photoionization cross sections calculations for neutral nitrogen and oxygen
showed excellent accord with high resolution measurements made at the Advanced Light Source (ALS)
radiation facility \cite{McLaughlin2011,Stolte2013} as have similar cross section calculations on singly and multiply 
ionized stages of atomic nitrogen compared with high resolution 
measurements at the SOLEIL synchrotron facility \cite{Soleil2011,Soleil2013}. 
 The majority of the high-resolution experimental data from third 
 generation light sources have been shown to be in excellent agreement with 
the state-of-the-art R-matrix method \cite{rmat,codes}  and with other modern theoretical approaches.

The present  investigation for this proto-type B-like atomic nitrogen ion gives accurate
values of photoionization cross sections produced by x-rays in the vicinity of the {\it K}-edge, 
where strong n=2 inner-shell resonance states of N$^{2+}$ are observed. 
This work compliments our previous studies on {\it K}-shell photoionization of singly and multiply ionized 
atomic nitrogen ions \cite{Soleil2011,Soleil2013} in the vicinity of the K-edge.
Previous experimental studies on B-like atomic nitrogen have been 
performed only in the valence shell region \cite{bizau2004,bizau2005}.
To date no experimental studies for the {\it K}-edge region have been reported  in the literature.  
For the  B-like ions N$^{2+}$, O$^{3+}$ and F$^{4+}$, absolute 
experimental cross section measurements in the near threshold region were 
made by  Bizau and co-workers \cite{bizau2004,bizau2005}. 
Close-coupling calculations performed by Li-Guo and Xin-Xiao \cite{china2009} 
based on the R-matrix formalism \cite{Burke2011} obtained excellent 
agreement with the experimental work of Bizau and co-workers \cite{bizau2005}.
In this near threshold region it was necessary to include both the ground state and metastable excited states 
in the theoretical work in order to achieve suitable agreement with experiment. 

 We follow a similar approach here for the {\it K}-shell energy region. N$^{2+}$ ions 
produced in the SOLEIL synchrotron radiation experiments are not purely in their ground state. 
{\it K}-shell photoionization contributes to the ionization balance in a
more complicated way than outer shell photoionization. In
fact {\it K}-shell photoionization when followed by Auger decay couples
three or more ionization stages instead of two in 
the usual equations of ionization equilibrium \cite{Petrini1997}.

The 1s $\rightarrow$ 2p photo-excitation process 
on the $\rm 1s^22s^22p~^2P^o$ ground-state of B-like nitrogen ion is,
$$
 h\nu + {\rm N^{2+}(1s^22s^22p~^2P^o)}  \rightarrow  {\rm N^{2+} ~ (1s2s^2 \,2p^2[^3P, ^1D, ^1S] ~^2S, ^2P, ^2D) }
 $$
 which can decay to
 $$
 \swarrow \quad \searrow
 $$
 $$
{\rm  N^{3+}~ (1s^22s^2~^1S) + e^- ({\it k^2_{\ell}}),} \; \; {\rm or} \; \;
{\rm  N^{3+}~ (1s^22s~ np~^{1}P^{\circ}) + e^- ({\it k^2_{\ell}}),}
$$
where $k^2_{\ell}$ is the outgoing energy of the continuum electron with angular momentum $\ell$.
Experimental studies on this doubly ionized atomic nitrogen ion, in its ground state
$\rm 1s^22s^22p~^2P^o$, are also hampered by metastable states present in the parent ion beam.  
In the present experimental studies performed at the SOLEIL radiation facility, 
N$^{2+}$ ions are produced in the gas-phase using an Electron-Cyclotron-Resonance-Ion-Source (ECRIS) 
so the metastable state $\rm 1s^22s2p^2~^4P$  can be present 
in the parent ion beam.  

For the $\rm 1s^22s2p^2~^4P$ metastable state, auto-ionization processes occurring 
by the 1s $\rightarrow$ 2p photo-excitation process  are mainly;
$$
 h\nu + {\rm N^{2+}(1s^22s2p^2~^4P)}
 $$
$$
\downarrow
$$
$$
{\rm  N^{2+} [1s2s[^{1,3}S]\,2p^3(^4S^o,^2D^o,^2P^o)]^4S^o,^4P^o,^4D^o}
$$
$$
\downarrow
$$
$$
{\rm N^{3+} (1s^22s~np~ ^3P^o) + e^- ({\it k^2_{\ell}}).}
$$
In the vicinity of the {\it K}-edge our current investigations appear to 
be the first time this B-like system has been studied experimentally for such an energy region.

State-of-the-art  {\it ab initio} calculations for Auger inner-shell processes were first carried out on this B-like system by 
Petrini and de Ara\'ujo \cite{Petrini1997} using the R-matrix method \cite{rmat} and followed a 
similar procedure to the work on {\it K}-shell studies for the Be-like B$^+$ ion \cite{Petrini1981}. 
Stoica and co-workers \cite{Petrini1998} noted that once  the 1s-hole was created in the ions, by single photoionization, with
simultaneous shake-up and shake-off processes, Auger decay
populates directly excited states of the residual ions, which then produces UV lines.
This work was further extended by  Garcia and co-workers \cite{Witthoeft2009}, using the optical potential
method within the Breit-Pauli R-matrix formalism \cite{rmat,codes,damp,Burke2011},
for photoionization of the ground state only, along the nitrogen iso-nuclear sequence.
Garcia and co-workers  \cite{Witthoeft2009}  pointed out that
 the earlier central field calculations \cite{Verner1993,Verner1995,rm1979} did 
 not account for the strong autoionizing resonance features 
 that dominate the cross sections near the {\it K}-edge.

In the present study we compare our  theoretical cross section 
results from the multi-configuration Dirac Fock (MCDF) and R-matrix methods 
with previous theoretical results \cite{Chen1987,Chen1988,Witthoeft2009} and with 
the current experimental measurements made at the SOLEIL synchrotron radiation facility.
Detailed measurements of the  {\it K}-shell  single photon ionization  cross sections 
for B-like nitrogen ions, were made in the 404 eV -- 409 eV region (where strong peaks were observed)
and in the photon energy range 439 eV -- 442 eV.  
The results for resonance energies and Auger widths are compared with detailed theoretical predictions made using 
the MCDF \cite{Bruneau1984}, R-matrix \cite{Burke2011} and the
SCUNC empirical methods \cite{Sakho2013a,Sakho2013b}.
The theoretical predictions assist in the identification and characterization 
of the strong $\rm 1s \rightarrow 2p$ and  the weaker $\rm 1s \rightarrow 3p$ 
resonances observed in the B-like nitrogen spectra. 
The current study gives absolute values (experimental and theoretical)
for PI cross sections along with n=2 and n=3 inner-shell resonance energies, natural line widths 
and resonance strengths, for the situation of a photon interacting with the ground 
$\rm 1s^22s^22p~^2P^o$,  and metastable $\rm 1s^22s2p^2~^4P$ states of the N$^{2+}$ ion.

In section 2 we briefly outline the experimental procedure used and
section 3 presents the theoretical procedures used. 
Section 4 gives a discussion of our experimental and theoretical results.
Finally in section 5 conclusions are drawn from the present study.

\section{Experiment}\label{sec:exp}

Cross sections for photoionization of B-like atomic nitrogen 
ions were measured in the range where {\it K}-shell photoionization
occurs. The experiment was performed at the MAIA (Multi-Analysis Ion Apparatus)
set-up, permanently installed on branch A of the PLEIADES beam line  \cite{Pleiades2010,Miron2013} at SOLEIL,
the French National Synchrotron Radiation Facility, located in Saint--Aubin,
France. Further details of the experimental setup were outlined in our previous 
publications on N$^{+}$ \cite{Soleil2011}, N$^{3+}$ and N$^{4+}$ \cite{Soleil2013} 
therefore only a brief summary will be given here.
The N$^{2+}$  ions are produced in a permanent magnet
electron cyclotron resonance ion source (ECRIS). 
Collimated N$^{2+}$  ion-beam currents
up to 160 nA were extracted from the ion source after biasing the ion source by
+2 kV and then selected by mass per charge ratio using a dipole magnet selector.
The ion beam was placed on the same axis as the photon beam by using
electrostatic deflectors and einzel lenses to focus the beam. 
After the interaction region between the photon and the ion beams, another dipole magnet
separates the primary beam and the beam of ions which have gained one 
(or several) charge(s) in the interaction, the so-called photo-ions. The
primary ions are collected in a Faraday cup and the photon-ions are detected by
multi-channel plates detector. The photon current is measured by a calibrated
photodiode. Ions with the same charge as the photo-ions can also 
be produced by collisions between the primary
ions and the residual gas or stripping on the walls in the interaction region.
This background signal is subtracted by chopping the photon beam, collecting the
data with and without photons for 20 seconds accumulation time.

For the absolute measurements of the PI cross sections, a -1000 V bias is applied
on the 50 cm long interaction-region and the data are collected with 30 meV
photon energy steps. The overlap of the two beams and the density distributions
of the interacted particles is determined in three dimensions by using three sets
of scanning slits.  The cross sections obtained have an estimated systematic uncertainty of 15\%. 
In another spectroscopy mode, no bias is applied to the
interaction region allowing the photon and ion beams to interact over about 1 m
and to scan the photon energy with a finer step. 
In this mode, only relative cross sections can be measured. 
They are later normalized on the cross sections determined 
in the absolute mode assuming the area under the resonances to be the same.
The energy and band width of the photon beam are 
calibrated separately using a gas cell and N$_{2}$ (1s $\rightarrow{}$
$\pi{}$*$_{g}$  v=0) photoionization lines, located at 400.87 eV \cite{Sodhi1984} and 
Ar 2p$_{3/2}$$^{-1}$4s at 244.39 eV \cite{Kato2007}. The photon energy, 
once corrected for Doppler shift, has an uncertainty of approximately 30 meV. 
The relative uncertainty on the band passes is of the order of 10 \%.
 Outstanding possibilities in terms of spectral resolution and flux at 
 the N$_2$ (1s$^{-1}$) $K$-edge have been discussed 
recently by Miron and co-workers \cite{Miron2012,Kimberg2013}.
\section{Theory}\label{sec:Theory}

\subsection{SCUNC: B-like nitrogen}\label{subsec:SU_Theory}
In the framework of the Screening Constant by Unit Nuclear Charge (SCUNC) 
formalism \cite{Sakho2013a,Sakho2013b}, the total energy
of the core-excited states is expressed in the form given by,
\begin{equation}
E  \left(  N\ell n\ell^{\prime}; ^{2S+1}L^{\pi}   \right ) 
= -Z^2  \left[  \frac{1}{N^2} 
+ \frac{1}{n^2} \left[  1 -\beta \left (  N\ell n\ell^{\prime}; ^{2S+1}L^{\pi}; Z  \right )  \right ]^2  \right ] 
\end{equation}
where $E \left( N\ell n\ell^{\prime}; ^{2S+1}L^{\pi}  \right )$ is in Rydberg units. 
In this equation, the principal quantum numbers $N$ and $n$ are respectively for the inner and the
outer electron of the He-like iso-electronic series. The $\beta$-parameters are screening constants by
unit nuclear charge expanded in inverse powers of $Z$ and are given by the expression,
\begin{equation}
 \beta \left( N\ell n\ell^{\prime}; ^{2S+1}L^{\pi}  \right )  = \sum_{k=1}^{~} f_k \left( \frac{1}{Z}  \right )^k 
\end{equation}
where $f_k  \left(  N\ell n\ell^{\prime}; ^{2S+1}L^{\pi}  \right )$ are parameters 
that are evaluated empirically from existing experimental measurements on resonance energies.  
Similarly one may get the Auger widths $\Gamma$ in Rydbergs (1 Rydberg = 13.605698 eV)  from the formula
\begin{equation}
\Gamma ({\rm Ry})  = Z^2  \left [  1 - \sum_{q}^{~}  f_q ~ \left (  \frac{1}{Z}   \right ) ^q   \right ] ^2 .
\end{equation}
The Advanced Light Source experimental measurements of Schlachter and co-workers on K-shell photoionization  
of B-like carbon ions \cite{Schlachter2004} 
were used to determine the appropriate empirical parameters $f_q$ for the Auger widths. 

\begin{figure}
\begin{center}
\includegraphics[scale=2.5,height=18.0cm,width=16.0cm]{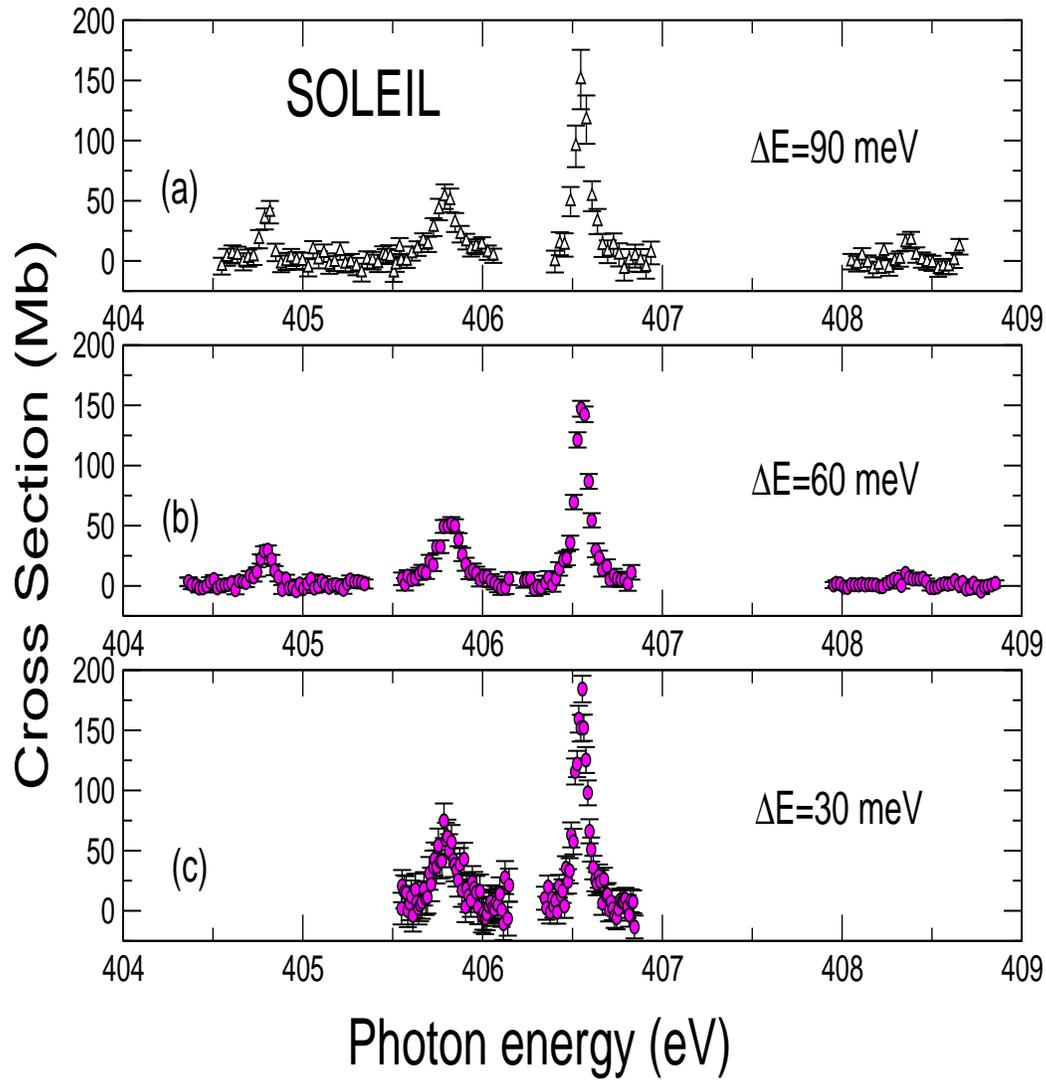}
\caption{\label{fig:90meV}(Colour online) Experimental K-shell photoionization cross sections for N$^{2+}$ ions measured 
					with various band passes at the SOLEIL synchrotron radiation facility 
					over the photon energy region 404 eV - 409 eV.
					(a) absolute results taken with a band pass of 90 meV FWHM, 
					obtained in the relative mode then normalized to the scan (a).
					For scan (a), the error bars give the total uncertainty, 
					and the statistical uncertainty for scans (b) and (c).}
\end{center}
\end{figure}

\subsection{MCDF: B-like Nitrogen}\label{subsec:MCDF_Theory}
Multi-configuration Dirac-Fock (MCDF) calculations were performed based on a full intermediate 
coupling regime in a $jj$-basis using the code developed by Bruneau  \cite{Bruneau1984}. Photoexcitation cross 
sections have been carried out for  B-like atomic nitrogen ions in the region of the
K-edge. Only electric dipole transitions have been computed using the length form.
For  this B-like atomic nitrogen ion the following initial configurations were considered:
$\rm {\overline{1s}}^2 {\overline{2s}}^2 {\overline{2p}}$ and $\rm {\overline{1s}}^2 {\overline{2s}}\; {\overline{2p}}^2$. 
In order to take account of correlation and relaxation effects, multiple orbital 
with the same quantum number were used.  The bar over the orbital indicates that they are different for the initial and final states.
In order to describe the correlation and relaxation effects, multiple orbitals with the same 
quantum number and the following final configurations  were considered:
 $\rm 1s 2s^2 2p^2$, $\rm 1s 2s 2p^2 np$, $\rm 1s 2s 2p^3$, $\rm 1s 2s^2 2pnp$.
Such notation means that radial functions with principal quantum number n =1 or 2 are not the same for initial 
and final configurations. Radial functions with principal quantum number up to n=6 were included in our calculations.
Photoexcitation cross sections from the $\rm 1s^2 2s^2 2p$ and  $\rm 1s^2 2s 2p^2$ configurations 
were calculated separately.
The wavefunctions were calculated minimizing the following energy functional:
\begin{equation}
E =\frac{ \sum_{\alpha}(2J_{\alpha} + 1) E_{\alpha}}{2 \sum_{\alpha}(2J_{\alpha} + 1)}    + 
       \frac{ \sum_{\beta}(2J_{\beta} + 1) E_{\beta}}{2 \sum_{\beta}(2J_{\beta} + 1)} 
\end{equation}
where $\alpha$ and $\beta$ run over all the initial and final states, respectively. 

Synthetic spectra were constructed as a weighted sum of the photoexcitation cross sections 
from the ground state levels  $\rm 1s^2 2s^2 2p(^2P_{1/2,3/2})$ and the metastable state levels 
$\rm 1s^2 2s 2p^2 (^4P^{\circ}_{1/2,3/2,5/2})$.  In order to compare with the SOLEIL experiments 
the weights used were  80 \% of the $\rm ^2P^{\circ}$ levels, and 20 \% for the $\rm ^4P$.
Each electric dipole transition has been dressed by a Lorenztian profile with a full width half maximum (FWHM)
given by the autoionization width of the upper level.  The autoionization widths were evaluated from our MCDF calculations
and for the photo-excited configurations $\rm 1s 2s^2 2p^2$ and  $\rm 1s 2s 2p^3$.
The average Auger widths were found respectively to be 64 meV and 105 meV.

\begin{figure}
\begin{center}
\includegraphics[scale=2.5,height=14.0cm,width=15.0cm]{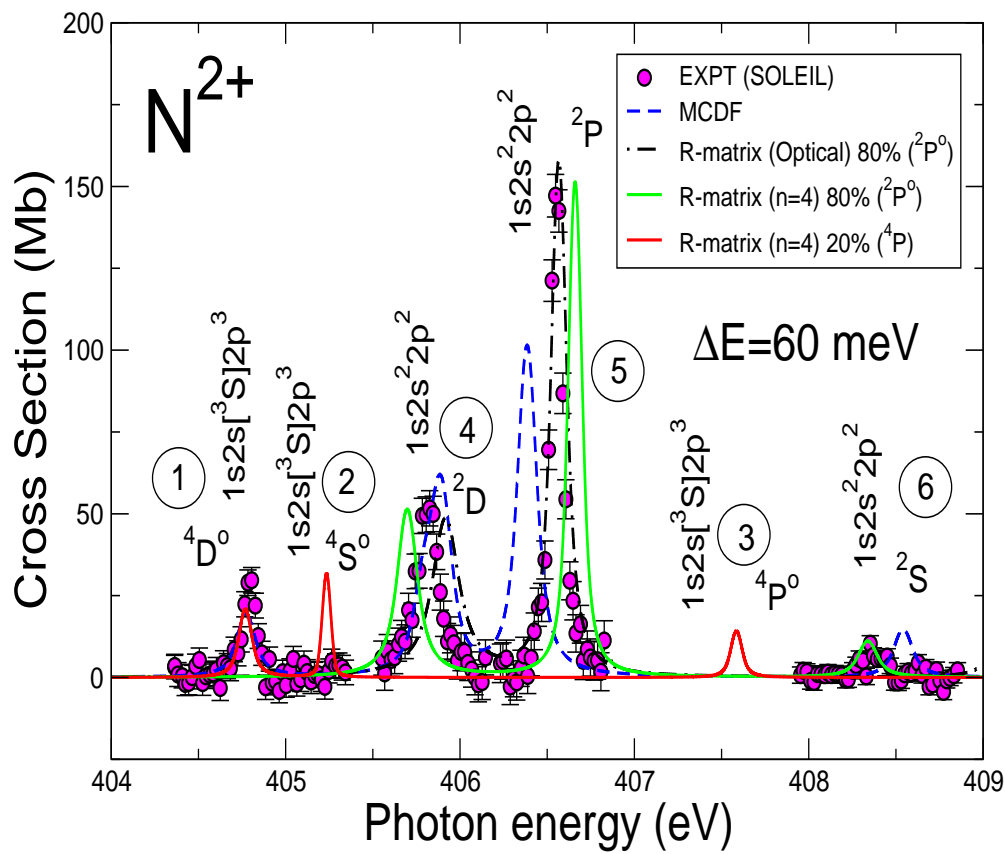}
\caption{\label{fig:60meV}(Colour online) Photoionization cross sections for N$^{2+}$ ions measured 
								with a 60 meV band pass at SOLEIL. 
								Solid circles : total photoionization. 
								The error bars give the statistical uncertainty of the experimental data. 
								The MCDF (dashed line), R-matrix RMPS
								 (solid line, red,  $\rm ^4P$, green, $\rm ^2P^o$), 
								 and the Optical potential (dash-dot line, $\rm ^2P^o$ only)  \cite{Witthoeft2009}
								calculations shown were obtained  by convolution with a Gaussian distribution
								 having a profile width at FWHM of 60 meV 
								 and a weighting of the ground and metastable states 
								 (see text for details) to simulate the measurements.  
								Tables 1 and 2 gives the designation and the parameters 
								 of the resonances~\mycirc{1}  - \mycirc{6}
								 in this photon energy region.}
\end{center}
\end{figure}

\begin{figure}
\begin{center}
\includegraphics[scale=1.5,height=14.0cm,width=16.0cm]{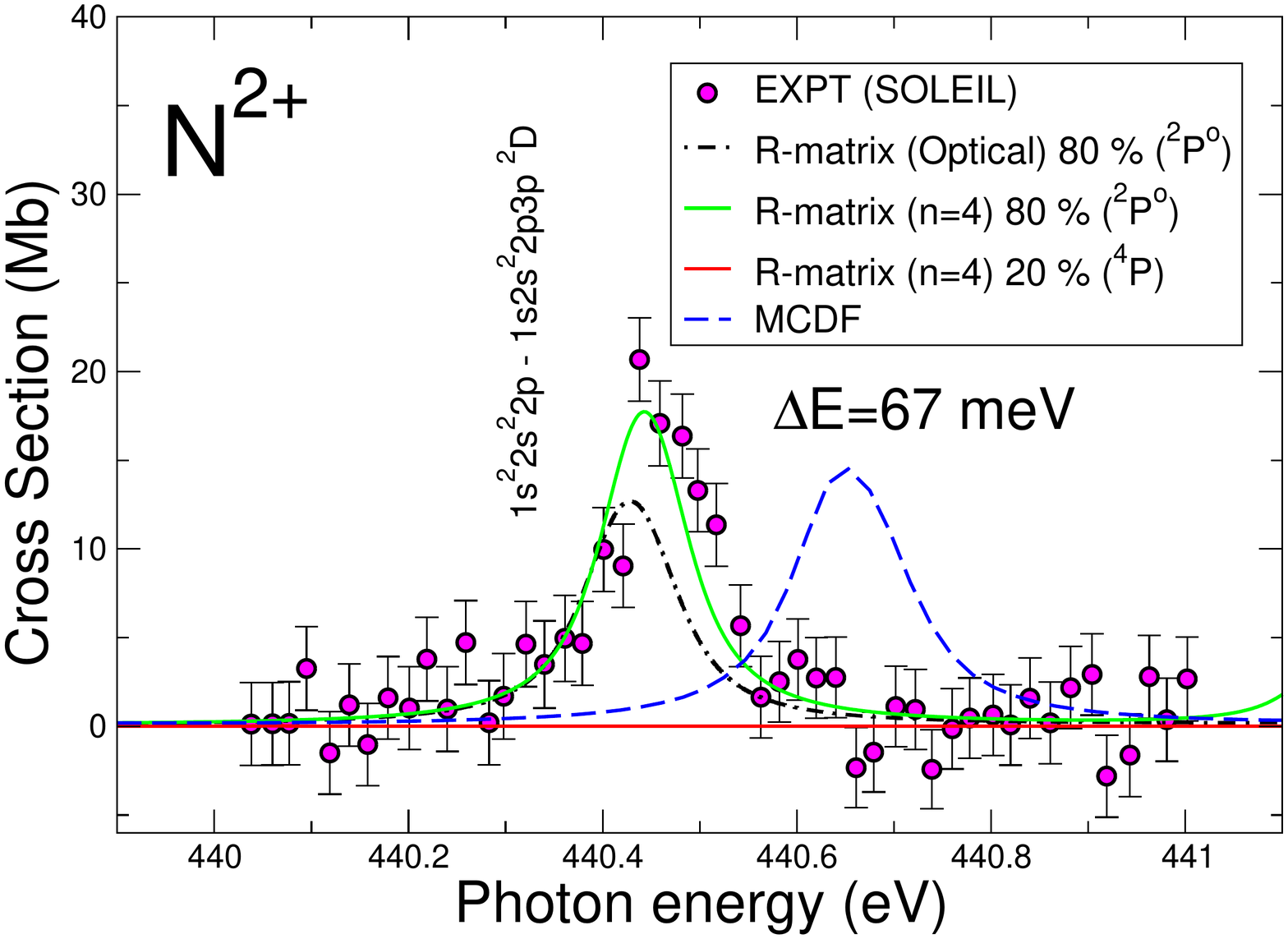}
\caption{\label{fig:67meV}(Colour online) Photoionization cross sections for N$^{2+}$ ions measured 
								with a 67 meV FWHM band pass at the SOLEIL radiation facility. 
								Solid circles : total cross sections. 
								The error bars give the statistical uncertainty of the experimental data. 
								 The MCDF (dashed line, blue), 
								 R-matrix  RMPS (solid line, red $\rm ^4P$, green $\rm ^2P^o$) 
								 and  the Optical potential (dash-dot line, $\rm ^2P^o$ only) \cite{Witthoeft2009}
								calculations shown were obtained  by convolution with a Gaussian
								 distribution having a profile width at FWHM of 67 meV and 
								 a weighting of the ground and metastable states 
								 (see text for details) to simulate the measurements.}
\end{center}
\end{figure}

\subsection{R-matrix: B-like Nitrogen}\label{subsec:R-Matrix_Theory}
The $R$-matrix method \cite{Burke2011}, implemented in a parallel version 
of the codes \cite{ballance06,McLaughlin2012,Ballance2012} was used to determine
all the cross sections.  The PI cross sections were carried out for the initial $\rm ^2P^o$ ground state 
and the  $\rm ^4$P metastable states.  All the PI cross section calculations were performed 
in $LS$-coupling with 390-levels  of the residual ion retained in the close-coupling expansion.
The  $\rm 1s$, $\rm 2s$ and $\rm 2p$ tabulated orbitals from the work of Clementi and Roetti
\cite{Clementi1974} were used together with n=3 physical and n=4 pseudo-orbitals 
of the N$^{3+}$ residual ion. The n=4 pseudo-orbitals were energy optimized on the 
 appropriate $\rm 1s^{-1}$ hole-shell states \cite{Berrington1997}, employing the atomic structure code CIV3 \cite{Hibbert1975}.  
These n=4  so called pseudo-orbitals are incorporated into the  scattering basis set  to try and accommodate 
for core relaxation and electron correlation effects in the multi-reference-configuration interaction wavefunctions 
used to describe the atomic ion states. All the N$^\mathrm{3+}$ residual ion states were then represented
by using multi-reference-configuration-interaction (MRCI) wave functions. The non-relativistic
$R$-matrix approach was used to calculate the energies
of the N${^\mathrm{2+}}$ bound states and the subsequent PI cross sections.
 
 The R-matrix with pseudo-states method (RMPS) was used  to determine all the cross sections (in $LS$ - coupling) with
390 levels of the N$^\mathrm{3+}$ residual ion included in the  close-coupling calculations.   
Since metastable states are present in the parent ion beam,  theoretical PI cross-section calculations are required 
for both the $\rm 1s^22s^22p~^2P^o$ ground state and the  $\rm 1s^22s2p^2~^4$P metastable 
states of the N$^\mathrm{2+}$ ion for a proper comparison with experiment. 

The scattering wavefunctions were generated by allowing two-electron promotions out of selected base
configurations of N$^\mathrm{2+}$. Scattering calculations were performed with twenty
continuum functions and a boundary radius of 9.4 Bohr radii. For the $\rm ^2$P$^o$ ground state and the  $\rm ^4P$
metastable states the electron-ion collision problem was solved with a fine energy mesh of 
2$\times$10$^{-7}$ Rydbergs ($\approx$ 2.72 $\mu$eV) to delineate all the resonance features in the PI cross sections.  
Radiation and Auger damping were also included in the cross section calculations.  

For a direct comparison with the SOLEIL results,  the R-matrix cross section calculations 
were convoluted with a Gaussian function of appropriate width and 
an admixture of 80\% ground  and 20 \% metastable states used to best  simulate experiment. 

The peaks  found in  the theoretical photoionization cross section 
spectrum were fitted to Fano profiles for overlapping resonances 
\cite{Fano1968,Shore1967,Morgan2008}
as opposed to the energy derivative of the eigenphase sum method \cite{keith1996,keith1998,keith1999}. 
The theoretical values for the natural linewidths $\Gamma$ determined from this procedure  
are presented in Tables 1 and 2 and compared with results obtained from 
the high-resolution SOLEIL synchrotron measurements and with other theoretical methods.

R-matrix calculations were also performed using an n=2 basis (1s, 2s and 2p) comparable
to the work of Garcia and co-workers \cite{Witthoeft2009} for the $\rm ^4P$ 
metastable.  The appropriate resonance parameters (for the $\rm ^4P$ metastable 
and the $\rm ^2P^o$ ground state similar to the work of Garcia and co-workers  \cite{Witthoeft2009})
are included in Tables \ref{reson} and \ref{reson2} from these calculations for completeness.

\section{Results and Discussion}\label{sec:Results}
Over the two photon energy ranges investigated 
we found  in the cross sections  that the most intense resonance peaks were 
located in the photon energy region, 404 eV -- 409 eV. 
All the experimental measurements are shown in figure 1, taken at the  various 
spectral  photon energy resolutions, ranging from 30 meV to 90 meV at FWHM.
We note that since the $^4$P metastable state of the N$^{2+}$  ion is present in the parent
experimental beam, theory may be used to estimate its content. 
Figure 2 shows our experimental and theoretical (MCDF and R-matrix) results 
for the restrictive photon energy range of 404 eV -- 409 eV.  
From our theoretical R-matrix studies on this system we find that an  admixture 
of 80\% ground state and 20 \% metastable state in 
the parent ion beam indicates satisfactory agreement between theory and experiment.
This admixture was determined by comparing the area under the strong peaks between 
experiment and theory.
 %
%
%
%
%
\begin{table}
\caption{\label{reson} B-like atomic nitrogen ions, quartet core-excited states. Comparison of the 
				  present experimental and theoretical results for the resonance energies $E_{\rm ph}^{\rm (res)}$ (eV),
            			  natural linewidths $\Gamma$ (meV) and resonance strengths $\overline{\sigma}^{\rm PI}$ (in Mb eV),
           			  for the dominant core photoexcited n=2 states of the N$^{2+}$ ion, in the photon energy region 
           			  404 eV to  409 eV with previous investigations.} 
 \lineup
  \begin{tabular}{ccr@{\,}c@{\,}llcl}
\br
 Resonance    & & \multicolumn{3}{c}{SOLEIL}                                       			& \multicolumn{1}{c}{R-matrix} 		& \multicolumn{2}{c}{MCDF/Others}\\
 (Label)            & & \multicolumn{3}{c}{(Experiment$^{\dagger}$)}      				& \multicolumn{1}{c}{(Theory)} 		& \multicolumn{2}{c}{(Theory)}\\
 \ns
 \mr
 \\
 $\rm 1s2s[^3S]2p^3\, ^4D^{\circ}$				
                                  & $E_{\rm ph}^{\rm (res)}$     	&	 	& 404.794 $\pm$ 0.03$^{\dagger}$ 	& 				&404.776$^{a}$ &	& 404.784$^{b}$  \\
  ~ \mycirc{1}	         &           				        	&		&  						  	&				&404.623$^{e}$ &	& 405.630$^{c}$  \\
				&						&		&							&				&			 &	& 404.930$^{d}$  \\
 			         \\
			  	& $\Gamma$ 			       	& 	   	 & 56 $\pm$ 19  	     			& 	         			&  62$^{a}$ 	&	&   			\\
 	    			&           				       	&   		 &  						         &				&  68$^{e}$   	&	&  \031$^{c}$	\\ 
				&					       	&		 &							&				&			&	&  \056$^{d}$	\\	  
				& $\overline{\sigma}^{\rm PI}$ &		& 3.21 $\pm$ 0.74				&				& 2.53$^{a}$	&	&  \0--		 \\
\\
$\rm 1s2s[^3S]2p^3\, ^4S^{\circ}$		 		
   				& $E_{\rm ph}^{\rm (res)}$      	&   		&   --							& 				&405.234$^{a}$ &	& 405.780$^{b}$ \\
  ~\mycirc{2}		&          	 		               	&   		&  				 		         &				&405.553$^{e}$ &	& 404.354$^{c}$  \\
   				&					      	&		&							&				&			&	& 405.510$^{d}$   \\ 		 
				&          	 		               	&   		&  				 		         &				&     			&	& 			 \\
				\\
                                     & $\Gamma$                        	&		&    --            		        			&     				&  \015$^{a}$  	 &	&   --     		\\
		 	         &                                              	&   		&  						         &				&  \016$^{e}$    &	&  \012$^{c}$     \\
			         &					      	&		&							&				&			&	&  \011$^{d}$	 \\
			         \\
				&$\overline{\sigma}^{\rm PI}$ 	&		&  --							&				& \02.19$^{a}$	&	& \0--		 \\
 	   		       \\
$\rm 1s2s[^3S]2p^3\, ^4P^{\rm o}$				
     				& $E_{\rm ph}^{\rm (res)}$    	& 		& -							&  				&407.584$^{a}$&	& 408.543$^{b}$ \\
 ~\mycirc{3} 	         &           				    	&   		&  							&    			 	&408.273$^{e}$ &    	& 408.411$^{c}$  \\
 				&					    	&		&							&				&			&	& 407.580$^{d}$ \\
			         \\
				& $\Gamma$			    	&    		& --				     			&  		     		& \048$^{a}$  	 &	&  \0--      		\\
 	    			&           				    	&   		&  							&				& \051$^{e}$   	&	& \024$^{c}$       \\ 
				&					    	&		&							&				&			&	& \015$^{d}$	   \\  
				\\
				&$\overline{\sigma}^{\rm PI}$ 	&		&  --							&				& \01.35$^{a}$	&	& \0--	     \\
				 \\           
  \br
\end{tabular}
~\\
$^{\dagger}$SOLEIL, experimental work.\\
$^{a}$R-matrix, n=4 basis,  $LS$-coupling,  present work.\\
$^{b}$MCDF, present work \\
$^{c}$MCDF, Chen and co-workers. \cite{Chen1987,Chen1988}\\
$^{d}$Screening Constant by Unit Nuclear Charge (SCUNC) approximation, present work.\\
$^{e}$R-matrix, n=2 basis, intermediate coupling, level averaged.\\
\end{table}

%
%
%
%
%
\begin{table}
\caption{\label{reson2} B-like atomic nitrogen ions, doublet core-excited states. Comparison of the present experimental
				    and theoretical results for the resonance energies $E_{\rm ph}^{\rm (res)}$ (eV),
            			    natural linewidths $\Gamma$ (meV) and resonance strengths $\overline{\sigma}^{\rm PI}$ (in Mb eV),
           			   for the dominant core photoexcited n=2 states of the N$^{2+}$ ion, in the photon energy region 
           			   404 eV to  409 eV with previous investigations.} 
 \lineup
  \begin{tabular}{ccr@{\,}c@{\,}llcl}
\br
 Resonance    & & \multicolumn{3}{c}{SOLEIL}                                       & \multicolumn{1}{c}{R-matrix} 		& \multicolumn{2}{c}{MCDF/Others}\\
 (Label)            & & \multicolumn{3}{c}{(Experiment$^{\dagger}$)}      & \multicolumn{1}{c}{(Theory)} 		& \multicolumn{2}{c}{(Theory)}\\
 \ns
 \mr
\\
 $\rm 1s2s^22p^2\, ^2$D				
  				& $E_{\rm ph}^{\rm (res)}$   & 		         & 405.814 $\pm$ 0.03$^{\dagger}$	 & 				& 405.703$^{a}$       & & 405.890$^{b}$  \\
  ~\mycirc{4} 		&           				   &   	 		&  						          &				& 405.965$^{e}$	& & 406.128$^{c}$  \\
   		   		 &          				   &		   	&  						          &				&   --  			& & 405.980$^{d}$	   \\
   		   		 &          				   &		   	&  						          &				&  --   			& &	   \\
				 \\
				& $\Gamma$ 			  & \;\0\0\    	& 122 $\pm$ 19    				& 		   		&122$^{a}$       	& &\055$^{c}$ \\
  	    			&          				  &   			&  							&				&109$^{e}$      	        & & 123$^{d}$       \\  
				    \\  
				&$\overline{\sigma}^{\rm PI}$&			&  10.1 $\pm$ 1.9				&				&10.44$^{a}$		& &\0--		    \\
	    			\\            
 $\rm 1s2s^22p^2\, ^2P$				
  				& $E_{\rm ph}^{\rm (res)}$ & 			& 406.547 $\pm$ 0.03$^{\dagger}$	&  				&406.656$^{a}$    	& & 406.380$^{b}$    \\
  ~\mycirc{5} 		&           			          &   	 		&  						         &				&406.387$^{e}$	& & 406.404$^{c}$  \\
 	    			&           				 &   			&  							&				&     				& & 406.561$^{d}$	     \\
				\\
				& $\Gamma$			 & \;\0\0\0    	& 58 $\pm$  7   				&  		     		& \0\062$^{a}$   	& &  \\
 	    			&          				 &   			&  							&				&  \0\043$^{e}$   	& & \025$^{c}$   \\              
 	    			&          				 &   			&  							&				& 		 	        & & \066$^{d}$   \\              
				\\
				&$\overline{\sigma}^{\rm PI}$&			& 16.80 $\pm$ 2.7				&				&19.32$^{a}$		& &\0--			    \\
 	   		      \\
 $\rm 1s2s^22p^2\, ^2$S				
   				& $E_{\rm ph}^{\rm (res)}$ & 		         &  408.376 $\pm$ 0.03$^{\dagger}$	& 				&408.344$^{a}$    	& & 410.085$^{b}$ \\
  ~\mycirc{6} 		&           				&   			&  						        	&				&408.297$^{e}$	& & 408.087$^{c}$ \\	
   		  		 &           				&   	 		&  						         &				&     				& & 408.414$^{d}$   \\
   		  		 &           				&   	 		&  						         &				&     				& &\0--   \\
				 \\
 				& $\Gamma$			& \;\0\0\    		& 120  $\pm$ 60				&    				& \0106$^{a}$  		& &  \025$^{c}$    	   \\
	    			&          				&   			&  							&				& \0\094$^{e}$    	& & 132$^{d}$  \\  
				\\            
				&$\overline{\sigma}^{\rm PI}$&			&  1.18 $\pm$ 0.47				&				& \01.87$^{a}$		& &\0--			    \\
				 \\           
  \br
\end{tabular}
~\\
$^{\dagger}$SOLEIL, experimental work.\\
$^{a}$R-matrix, n=4 basis, $LS$-coupling,  present work.\\
$^{b}$MCDF, present work\\
$^{c}$MCDF, Chen and co-workers. \cite{Chen1987,Chen1988}\\
$^{d}$Screening Constant by Unit Nuclear Charge (SCUNC) approximation, present work.\\
$^{e}$R-matrix, n=2 basis, intermediate coupling, level averaged. \\
\end{table}

For a direct comparison with the experimental work both the MCDF and R-matrix theoretical cross sections 
were convoluted with a Gaussian distribution function having a profile width at 
FWHM of 60 meV and the above admixture used to simulate the SOLEIL measurements 
and the results presented in figure 2.  
Figure 2 illustrates the experimental and theoretical  (MCDF and R-matrix) single photonionization 
cross sections results over the photon energy range from 404 eV -- 409 eV at the energy resolution of 60 meV. 
Tables \ref{reson} and \ref{reson2} gives the experimental and theoretical estimates 
for all of the resonance parameters of the peaks located 
in the photon energy range 404 eV - 409 eV.  
From Table \ref{reson2}, one sees that the
MCDF predictions for the $^2$S resonance energy is at 410.1 eV, outside of the energy range investigated.
The peaks found in the photoionization cross section spectra measured  at SOLEIL for this
B-like atomic nitrogen were fitted to Voigt functions.
Voigt profiles were used to fit the peaks in the experimental data, in order to extract the line widths,
with a Gaussian instrumental profile of 60 meV assumed for each peak. For the determination 
of the experimental widths, each individual sweep has been fitted separately by  Voigt profiles 
to avoid any possible shift in the energy delivered by the monochromator. Then the final width of the 
Lorentzian component was obtained from the weighted mean of the individual Lorentzian width 
determined from each fit. For the N$^{2+}$ ion, splitting of the $J$ components of the initial state \cite{NIST2013} 
was also taken into account, assuming a statistical distribution of the levels. 
The experimental uncertainties in the tables are the total uncertainties.
 
The MCDF and R-matrix cross sections results indicate suitable agreement with experiment 
from matching the calculated and experimental ionization thresholds. 
Estimates for the resonance parameters, for the N$^{2+}$ doublet and quartet states, made using the 
screening constant by unit nuclear charge (SCUNC) empirical fitting 
approach \cite{Sakho2013a,Sakho2013b} 
can be seen to be in satisfactory agreement with the more sophisticated 
MCDF and R-matrix theoretical methods 
and experiment (c.f. Tables \ref{reson} and \ref{reson2}).  

For the peaks found in the cross sections in the photon energy range 404  eV - 409 eV,
the resonance energies and linewidths (c.f. Tables \ref{reson} and \ref{reson2}) all indicate 
 suitable agreement between theory and the available experimental measurements.  
 The n=2 intermediate doublet $\rm 1s2s^22p^2~^2S,~^2P,~^2D$ resonance states 
have a strong presence in the resulting photoionization  spectra 
in the photon energy range 404 eV -- 409 eV. 
The quartet $\rm 1s2s[^{3}S]2p^3~^4S^o,~^4P^o,~^4D^o$ resonances 
have a weaker presence in the cross sections over this same photon energy range 
as illustrated in figure \ref{fig:60meV}.
 The $\rm ^4S^{\circ}$ peak is predicted theoretically (c.f. Table \ref {reson}) 
 but could be obscured by the $\rm ^4D^o$ resonance in the experimental spectra. 
 The two resonances are close in experimental energies and may not be 
 individually resolved (with the present resolution), 
 or the $\rm ^4S^{\circ}$ resonance could be much weaker than predicted. 
 Tentatively there appears some experimental 
 indication of this weak peak around about 405.2 eV but higher statistics than that used in the  
 present experiment would be required to resolve this feature.
  Similarly, missing from the present experimental measurements  
 is the $\rm ^4P^{\circ}$ peak (c.f. Table \ref{reson}), which is
 observed weakly, in analogous ALS experiments on B-like carbon
  (C$^{+}$) by Schlachter and co-workers \cite{Schlachter2004}. 
For the N$^{2+}$ ion our theoretical predictions indicate that the  
 $\rm 1s^22s2p^2~^4P \rightarrow 1s2s[^3S]2p^3(^2P)~^4P^o$ resonance energy is located 
 at an energy of about 407.6 eV which is outside of the energy region scanned 
 in the present experimental measurements.
 
 For B-like nitrogen, from tables \ref{reson} and \ref{reson2}, we note that peak \mycirc{1} ($\rm 1s2s[^{3}S]2p^3~^4D^o$) 
 and  \mycirc{5} ($\rm 1s2s^22p^2~^2P$)  have comparable Auger widths of about 60 meV. 
 Peaks  \mycirc{4} ($\rm 1s2s^22p^2~^2D$)  and  \mycirc{6}  ($\rm 1s2s^22p^2~^2S$)  also have similar 
 widths but are about a factor of two larger being around 120 meV. On physical grounds we attribute 
 this difference due to different branching ratio's and decay mechanism's to the ground 
 and excited states of the residual N$^{3+}$ ion. An analogy may be drawn with the decay mechanism's 
 in higher charged B-like ions, from the recent saddle-point calculations of Sun and co-workers \cite{Sun2013} 
 where similar trends were found.
  
 Due to dipole selection and Hund rules, angular momentum coupling consideration 
 give two possible $\rm ^4S^o$ states from the $\rm 1s2s[^{1,3}S]2p^3(^4S^o)$ states
 as opposed to a single one from the $\rm 1s2s[^3S]2p^3(^2D^o)~^4D^o$ state. 
 Highly correlated R-matrix  \cite{Guo2007} and saddle-point \cite{Sun2011}  
 calculations on B-like carbon confirm the earlier findings on the ordering of the quartet states
 by Schlachter and co-workers \cite{Schlachter2004}. For the present B-like system 
 our RMPS calculations give a similar ordering of the 
 quartet states presented in Table \ref{reson} as in C$^{+}$ \cite{Schlachter2004,Guo2007,Sun2011}.
Figure 2, ilustrates all of the N$^{2+}$ doublet states predicted  by theory, which are observed in this
 spectral range.  Theoretical predictions from the MCDF and R-matrix methods 
show suitable agreement with experiment, apart from the location of the $\rm ^2S$ resonance 
in the MCDF case. For the doublet states, the R-matrix results indicate better agreement 
with experiment (for cross sections and resonance parameters c.f. Table \ref{reson2}) 
than other theoretical methods.

Figure 3 shows the photon energy range in the vicinity of 
the $\rm 1s^22s^22p~^2P^o \rightarrow 1s2s^22p[^3P]3p~^2D$ 
transition which occurs from the ground state of the N$^{2+}$ ion. 
The SOLEIL measurements in this energy region were taken with a spectral resolution of 67 meV. 
To compare directly with experiment the MCDF and R-matrix 
calculations were convoluted with a Gaussian distribution function having a profile width at FWHM
of 67 meV and weighted 80\% for the ground-state and 20 \% for the metastable.
From figure 3 it is seen that the resonance line intensities, from the present 
RMPS, R-matrix calculations compare more favourably with 
the SOLEIL measurements. The experimental energy
of this resonance was found at 440.46 $\pm$ 0.04 eV with an Auger width of 
90 $\pm$ 47  meV  and a strength of  3.05 $\pm$ 0.83 Mb~eV.  
For this same resonance, R-matrix predictions give an energy of
440.441 eV, a width of 90 meV and a strength of 2.79 Mb~eV. MCDF predictions give 
an energy of 440.652 eV with a strength of 2.91 Mb ~eV, while SCUNC estimates
give values of 440.580 eV and 73 meV respectively for the energy and the width.
It is seen that the present theoretical estimates for this resonance (position and width) 
are in harmony with the experimental findings. On physical grounds
 we note that the R-matrix with pseudo-states (RMPS) approach 
provide additional electron-correlation which greatly improves the residual ion target structure,
 the ionization potential of the system, resonance energy positions and 
 Auger widths found in the photoionization cross-sections compared to 
 other approaches as can be seen from  the results presented in Tables \ref{reson} and \ref{reson2}.

 We note that XMM-{\it Newton}  X-ray spectral observations of WR 1 stars indicate 
 the NIV (N$^{3+}$) K-edge absorption edge feature is at about 460 eV (26.954 {\AA})
\cite{Ignace2003}.  For the K-edge a value of 459 eV (27.011 {\AA}) was quoted 
by Daltabuit and Cox \cite{Cox1972}, obtained from the screening constant method of 
Slater \cite{Slater1930,Slater1960}, which tends to overestimate the threshold as
indicated by Bethe and Salpeter \cite{Bethe1957}. We point out that similar wavelength $\lambda$  differences
($\Delta \lambda$ ~$\sim$ 0.06 {\AA} ) occur between K-edge experiment, observation and theoretical estimates
for atomic oxygen and its ions \cite{Stolte2013,Gatuzz2013}.  Modern methods such as the saddle-point with 
complex rotation \cite{Lin2001} give a value of 457.445 eV (27.1036 {\AA}), relative to the NIII (N$^{2+}$) ground state for this 
$\rm 1s2s^2 2p~ ^3P^o$  hole state of NIV (N$^{3+}$) in close agreement with our R-matrix estimate of 457.441 eV (27.1038 {\AA}).
Similarly for the NIV (N$^{3+}$) $\rm 1s2s^2 2p~ ^1P^o$  hole state, an estimate of 461.416 eV (27.870 {\AA}) 
from the saddle point method \cite{Lin2002}, is in close agreement with our R-matrix values 
of, 461.359 eV (26.874 {\AA}, n=3 basis), 461.422 eV (26.868 {\AA}, n=4 basis), 
and 461.317 eV (26.876 {\AA}, n=2 basis), from the earlier work of Garcia and co-workers \cite{Witthoeft2009}.

\section{Conclusions}\label{sec:Conclusions}
{\it K}-shell photoionization of B-like nitrogen ions, N$^{2+}$, has been investigated 
using state-of-the-art experimental and theoretical methods.
High-resolution spectroscopy was able to be achieved with E/$\Delta$E = 13,500 
at the SOLEIL synchrotron radiation facility, in Saint-Aubin, France, covering the
photon energy ranges; 404 eV -- 409 eV and 439 eV -- 442 eV.  
Several strong peaks are found in the cross sections in the energy regions,  404 eV - 409 eV 
and  439 eV -- 442 eV which are identified as the 1s $\rightarrow$ 2p and  1s $\rightarrow$ 3p transitions  
 in the N$^{2+}$ {\it  K}-shell spectrum and assigned spectroscopically with
 their resonance parameters tabulated in Tables \ref{reson} and \ref{reson2}.
 For the observed peaks, suitable agreement is found between the present theoretical and
experimental results both on the photon-energy scale and on the absolute
 cross-section scale for this prototype B-like system.  
 The strength of the present study is the high resolution of  the spectra along with 
 theoretical predictions made using state-of-the-art MCDF, R-matrix with pseudo-states and empirical methods.  
The present results have been compared with high resolution  experimental measurements made at the SOLEIL 
 synchrotron radiation facility and with other theoretical methods so
 would be suitable to be incorporated into astrophysical modelling codes like 
 CLOUDY \cite{Ferland1998,Ferland2003}, XSTAR \cite{Kallman2001} 
 and AtomDB \cite{Foster2012} used to numerically
simulate the thermal and ionization structure of ionized astrophysical nebulae. 

\ack
The experimental measurements were performed on the 
PLEIADES beamline \cite{Pleiades2010,Miron2013}, at the SOLEIL Synchrotron 
radiation facility in Saint-Aubin, France.  The authors would like to thank the SOLEIL staff and, 
in particular those of the PLEIADES beam line for their helpful assistance,
C Miron,  Xiao-Jing Liu, E Roberts and C Nicolas.
B M McLaughlin acknowledges support from the US National Science Foundation through a grant to ITAMP
at the Harvard-Smithsonian Center for Astrophysics, the RTRA network {\it Triangle de le Physique} 
and a visiting research fellowship from Queen's University Belfast. 
I Sakho acknowledges the hospitality of the Universit\'{e} Paris-Sud and support 
from the University Assane Seck of Ziguinchor during a recent visit.
We thank Dr John C Raymond and Dr Randall K Smith at the Harvard-Smithsonian Center 
for Astrophysics for discussions on the astrophysical applications and Dr Javier Garcia 
for numerical values of the cross sections  \cite{Witthoeft2009}.
The computational work was carried out at the National Energy Research Scientific
Computing Center in Oakland, CA, USA, the  Kraken XT5 facility at the National Institute 
for Computational Science (NICS) in Knoxville, TN, USA
and at the High Performance Computing Center Stuttgart (HLRS) of the University of Stuttgart, Stuttgart, Germany. 
We thank Stefan Andersson from Cray Research for his advice and assistance with 
the implementation and optimization of the parallel R-matrix codes on the Cray-XE6 at HLRS.
The Kraken XT5 facility is a resource of the Extreme Science and Engineering Discovery Environment (XSEDE), 
which is supported by National Science Foundation grant number OCI-1053575.
%
%
%
%
\section*{References}
\bibliographystyle{iopart-num}
\bibliography{n2plus}

\providecommand{\newblock}{}
\begin{thebibliography}{10}
\expandafter\ifx\csname url\endcsname\relax
  \def\url#1{{\tt #1}}\fi
\expandafter\ifx\csname urlprefix\endcsname\relax\def\urlprefix{URL }\fi
\providecommand{\eprint}[2][]{\url{#2}}

\bibitem{McLaughlin2001}
{McLaughlin B M} 2001 {} {\em {Spectroscopic Challenges of Photoionized
  Plasma}\/} ({\em ASP Con$f$. Series\/} vol \textbf{247}) ed {Ferland, G and
  Savin D W} (San Francisco, CA: Astronomical Society of the Pacific) p~87

\bibitem{Brickhouse2010}
{Foster A, Smith R, Brickhouse N, Kallman T, Witthoeft M} 2010 {\em {Space Sci.
  Rev.}\/} {\bf \textbf{157}} 135

\bibitem{Kallman2010}
{Kallman T~R} 2010 {\em {Space Sci. Rev.}\/} {\bf \textbf{157}} 177

\bibitem{Quinet2011}
{Quinet P, Palmeri P, Mendoza C, Bautista M, Garcia J, Witthoeft M, Kallman
  T~R} 2011 {\em {J Elect. Spec. and Relat. Phenom.}\/} {\bf \textbf{184}} 170

\bibitem{Soleil2011}
{Gharaibeh M F, Bizau J M, Cubaynes D, Guilbaud S, Hassan N El, Shorman M M Al,
  Miron C, Nicolas C, Robert E, Blancard C and McLaughlin B M} 2011 {\em {J.
  Phys. B: At. Mol. Opt. Phys.}\/} {\bf \textbf{44}} 175208

\bibitem{Soleil2013}
{Shorman M M Al, Gharaibeh M F, Bizau J M, Cubaynes D, Guilbaud S, Hassan N El,
  Miron C, Nicolas C, Robert E, Sakho I, Blancard C and McLaughlin B M} 2013
  {\em {J. Phys. B: At. Mol. Opt. Phys.}\/} {\bf \textbf{46}} 195701

\bibitem{McLaughlin2013}
{McLaughlin B M and Ballance C P} 2013 {Photoionization, Fluorescence and
  Inner-shell Processes} {\em {McGraw-Hill Yearbook of Science and Technology
  2013}\/} ed {McGraw-Hill} (New York, USA: McGraw-Hill Inc) p 281

\bibitem{Olge2004}
{Olge P M {\it et al}} 2004 {\em {Astrophys. J}\/} {\bf \textbf{606}} 151

\bibitem{Cassinelli1981}
{Cassinelli J P {\it et al}} 1981 {\em {Astrophys. J}\/} {\bf \textbf{250}} 677

\bibitem{Dutuit2013}
{Dutuit O {\it et al}} 2013 {\em {Astrophys. J}\/} {\bf \textbf{204}} 20

\bibitem{Diaz2007}
{ P\'{e}rez-Montero E and D\'{õ}az A} 2007 {\em {Mon. Not. Roy. Astr. Soc.}\/}
  {\bf \textbf{377}} 1195

\bibitem{Ignace2003}
{Ignace R {\it et al}} 2003 {\em {Astron. Astrophys.}\/} {\bf \textbf{408}} 353

\bibitem{Bohigas2008}
{Bohigas J} 2005 {\em {Astrophys. J}\/} {\bf \textbf 674} 954

\bibitem{Scully2006}
{Scully S W J, {\'A}lvarez I, Cisneros C, Emmons E D, Gharaibeh M~F, Leitner D,
  Lubell M S, M\"{u}ller A, Phaneuf R A, P\"{u}ttner R, Schlachter A S,
  Schippers S, Ballance C P and McLaughlin B M} 2006 {\em {J. Phys. B: At. Mol.
  Opt. Phys.}\/} {\bf 39} 3957

\bibitem{Scully2007}
{Scully S W J, {\'A}lvarez I, Cisneros C, Emmons E D, Gharaibeh M~F, Leitner D,
  Lubell M S, M\"{u}ller A, Phaneuf R A, P\"{u}ttner R, Schlachter A S,
  Schippers S, Ballance C P and McLaughlin B M} 2007 {\em {J. Phys. Conf. Ser.
  }\/} {\bf \textbf{58}} 387

\bibitem{DPI2013}
{McLaughlin B M} 2013 {\em {J. Phys. B: At. Mol. Opt. Phys.}\/} {\bf
  \textbf{46}} 075204

\bibitem{Mueller2010}
{M{\"u}ller A, Schippers S, Phaneuf R~A, Scully S W J, Aguilar A, Cisneros C,
  Gharaibeh M~F, Schlachter A~S and McLaughlin B~M} 2010 {\em {J. Phys. B: At.
  Mol. Opt. Phys.}\/} {\bf 43} 135602

\bibitem{Mueller2009}
{M{\"u}ller A, Schippers S, Phaneuf R~A, Scully S W J, Aguilar A, Covington A
  M, {\'A}lvarez I, Cisneros C, Emmons E D, Gharaibeh M~F, Schlachter A~S,
  Hinojosa G and McLaughlin B~M} 2009 {\em {J. Phys. B: At. Mol. Opt. Phys.}\/}
  {\bf 42} 235602

\bibitem{Mueller2014}
{M{\"u}ller A, Schippers S, Phaneuf R~A, Scully S W J, Aguilar A, Cisneros C,
  Gharaibeh M~F, Schlachter A~S and McLaughlin B~M} 2014 {\em {J. Phys. B: At.
  Mol. Opt. Phys.}\/} {\bf ~} in preparation for publication

\bibitem{Scully2005}
{Scully S W J, Aguilar A, Emmons E D, Phaneuf R A, Halka M, Leitner D, Levin J
  C, Lubell M S, P\"{u}ttner R, Schlachter A S, Covington A M, Schippers S,
  M\"{u}ller A and McLaughlin B M} 2005 {\em {J. Phys. B: At. Mol. Opt.
  Phys.}\/} {\bf \textbf{38}} 1967

\bibitem{Schlachter2004}
{Schlachter A S, Sant'Anna M M, Covington A M, Aguilar A, Gharaibeh M~F, Emmons
  E D, Scully S W J, Phaneuf R A, Hinojosa G, {\'A}lvarez I, Cisneros C,
  M\"{u}ller A and McLaughlin B M} 2004 {\em {J. Phys. B: At. Mol. Opt.
  Phys.}\/} {\bf \textbf{37}} L103

\bibitem{Kawatsura2002}
{Kawatsura K {\it et al}} 2002 {\em {J. Phys. B: At. Mol. Opt. Phys. }\/} {\bf
  \textbf{35}} {\it 4147}

\bibitem{Yamaoka2001}
{Yamaoka H {\it et al}} 2001 {\em {Phys. Rev.}\/} {\bf \textbf{A~65}} {\it
  012709}

\bibitem{McLaughlin2011}
{Sant'Anna M~M, Schlachter A~S, \"{O}hrwall G, Stolte W~C, D W Lindle D~W and
  McLaughlin B~M } 2011 {\em {Phys. Rev. Letts.}\/} {\bf \textbf{107}} {\it
  033001}

\bibitem{Krause1994}
{Krause M O} 1994 {\em {Nucl. Instr. and Meth. in Phys. Res. B}\/} {\bf
  \textbf{87}} 178

\bibitem{Menzel1996}
{Menzel A, Benzaid S, Krause M, Caldwell C D, Hergenhahn U and Bissen M } 1996
  {\em {Phys. Rev. A}\/} {\bf \textbf{54}} R991

\bibitem{Stolte1997}
{Stolte W C, Samson J A R, Hemmers O, Hansen D, Whitfield S B and Lindle D W }
  1997 {\em {J. Phys. B: At. Mol. Opt. Phys.}\/} {\bf \textbf{30}} 4489

\bibitem{Stolte2013}
{McLaughlin B M, Ballance C P, Bown K P, Gardenghi D J and Stolte W C} 2013
  {\em {Astrophys. J}\/} {\bf \textbf{771}} L8

\bibitem{rmat}
{Burke P G and Berrington K A} 1993 {\em {Atomic and Molecular Processes: An
  R-matrix Approach}\/} (Bristol, UK: IOP Publishing)

\bibitem{codes}
{Berrington K A, Eissner W and Norrington P~H} 1995 {\em {Comput. Phys.
  Commun.}\/} {\bf \textbf{92}} 290
  \urlprefix\url{http://amdpp.phys.strath.ac.uk/APAP}

\bibitem{bizau2004}
{Bizau J M {\it et al}} 2004 {\em {Phys. Scr.}\/} {\bf \textbf{T~110}} 57

\bibitem{bizau2005}
{Bizau J M {\it et al}} 2005 {\em {Astron. Astrophys.}\/} {\bf \textbf{439}}
  387

\bibitem{china2009}
{Wang Guo-Li and Zhou Xiao-Xin} 2009 {\em {Chinese Phys. B}\/} {\bf
  \textbf{18}} 3833

\bibitem{Burke2011}
{Burke P G} 2011 {\em {R-Matrix Theory of Atomic Collisions: Application to
  Atomic, Molecular and Optical Processes}\/} (New York, USA: Springer)

\bibitem{Petrini1997}
{Petrini~D and de Ara\'ujo~ F~X} 1997 {\em {Astron. \& Astrophys.}\/} {\bf
  \textbf{326}} 870

\bibitem{Petrini1981}
{Petrini~D} 1981 {\em {J. Phys. B: At. Mol.}\/} {\bf \textbf{14}} 3839

\bibitem{Petrini1998}
{Stoica S, Petrini~D and Bely-Dubau F} 1998 {\em {Astron. \& Astrophys.}\/}
  {\bf \textbf{334}} L26

\bibitem{Witthoeft2009}
{Garcia J, Kallman T~R, Witthoeft M, Behar E, Mendoza C, Palmeri P, Quintet P,
  Bautista M and Klapisch M} 2009 {\em {Astrophys. J Supp. Ser.}\/} {\bf
  \textbf{185}} 477

\bibitem{damp}
{Robicheaux F, Gorczyca T W, Griffin D C, Pindzola M S and Badnell N R} 1995
  {\em {Phys. Rev. A}\/} {\bf 52} 1319

\bibitem{Verner1993}
{Verner D~A {\it et al}} 1993 {\em {At. Data Nucl. Data Tables}\/} {\bf
  \textbf{55}} 233

\bibitem{Verner1995}
{Verner D~A and Yakovlev D~G} 1995 {\em {Astron. Astrophys. Suppl.}\/} {\bf
  \textbf{109}} 125

\bibitem{rm1979}
{Reilman R and Manson S T} 1979 {\em {Astrophys. J. Suppl. Ser.}\/} {\bf
  \textbf{40}} 815

\bibitem{Chen1987}
{Chen~M~H and Crasemann~B} 1987 {\em {Phys. Rev A}\/} {\bf \textbf{35}} 4579

\bibitem{Chen1988}
{Chen~M~H and Crasemann~B} 1988 {\em {At. Dat. Nucl. Dat. Tables}\/} {\bf
  \textbf{38}} 381

\bibitem{Bruneau1984}
{Bruneau J} 1984 {\em {J. Phys. B: At. Mol. Phys.}\/} {\bf \textbf{17}} 3009

\bibitem{Sakho2013a}
{Sakho I {\it et al}} 2013 {\em {At. Data. Nuc. Data Tables}\/} {\bf
  \textbf{99}} 447

\bibitem{Sakho2013b}
{Sakho I {\it et al}} 2013 {\em {Phys. Scr.}\/} {\bf \textbf{88}} 035302

\bibitem{Pleiades2010}
{Travnikova O {\it et al}} 2010 {\em {Phys. Rev. Letts.}\/} {\bf \textbf{105}}
  233001

\bibitem{Miron2013}
{Miron C {\it et al} 2013}
  \urlprefix\url{http://www.synchrotron-soleil.fr/portal/page/portal/Recherche/LignesLumiere/PLEIADES}

\bibitem{Sodhi1984}
{Sodhi R~N~S and Brion C~E} 1977 {\em {J. Electron Spectrosc. Relat.
  Phenom.}\/} {\bf \textbf{34}} 363

\bibitem{Kato2007}
{Kato M {\it et al}} 2007 {\em {J. Electron Spectrosc. Relat. Phenom.}\/} {\bf
  \textbf{160}} 39

\bibitem{Miron2012}
{Miron C {\it et al}} 2012 {\em {Nature Phys.}\/} {\bf \textbf{8}} 135

\bibitem{Kimberg2013}
{Kimberg V {\it et al}} 2013 {\em {Phys. Rev. X}\/} {\bf \textbf{3}} 011017

\bibitem{ballance06}
{Ballance C P and Griffin D C} 2006 {\em {J. Phys. B: At. Mol. Opt. Phys.}\/}
  {\bf \textbf{39}} 3617

\bibitem{McLaughlin2012}
{McLaughlin B M and Ballance C P} 2012 {\em {J. Phys. B: At. Mol. Opt.
  Phys.}\/} {\bf \textbf{45}} 085701

\bibitem{Ballance2012}
{McLaughlin B M and Ballance C P} 2012 {\em {J. Phys. B: At. Mol. Opt.
  Phys.}\/} {\bf \textbf{45}} 095202

\bibitem{Clementi1974}
{Clementi E and Roetti C} 1974 {\em {At. Data Nucl. Data Tables}\/} {\bf
  \textbf{14}} 177

\bibitem{Berrington1997}
{Berrington K, Quigley L, and Zhang H L} 1997 {\em {J. Phys. B: At. Mol. \&
  Phys.}\/} {\bf \textbf{30}} 5409

\bibitem{Hibbert1975}
{Hibbert A} 1975 {\em {Comput. Phys. Commun.}\/} {\bf \textbf{9}} 141

\bibitem{Fano1968}
{Fano U and Cooper J W} 1968 {\em {Rev. Mod. Phys.}\/} {\bf \textbf{40}} 441

\bibitem{Shore1967}
{Shore B W} 1967 {\em {Rev. Mod. Phys.}\/} {\bf \textbf{39}} 439

\bibitem{Morgan2008}
{Wright J D {\it et al}} 2008 {\em {Phys. Rev. A}\/} {\bf \textbf{77}} 062512

\bibitem{keith1996}
{Quigley L and Berrington K~A} 1996 {\em {J. Phys. B: At. Mol. Phys.}\/} {\bf
  \textbf{29}} 4529

\bibitem{keith1998}
{Quigley L, Berrington K A and Pelan J} 1998 {\em {Comput. Phys. Commun.}\/}
  {\bf 114} 225

\bibitem{keith1999}
{Ballance C P, Berrington K A and McLaughlin B M} 1999 {\em {Phys. Rev. A}\/}
  {\bf \textbf{60}} R4217

\bibitem{NIST2013}
{Kramida A E, Ralchenko Y, Reader J, and NIST ASD Team (2013),} {NIST Atomic
  Spectra Database (version 5.1),} National Institute of Standards and
  Technology, Gaithersburg, MD, USA
  \urlprefix\url{http://physics.nist.gov/PhysRefData/ASD/levels_form.html}

\bibitem{Sun2013}
{Sun Y {\it et al}} 2013 {\em {Eur. Phys. J. D}\/} {\bf \textbf{67}} 88

\bibitem{Guo2007}
{Guo-Li W and Xiao-Xi Z} 2007 {\em {Chinese Phys.}\/} {\bf \textbf{16}} 2361

\bibitem{Sun2011}
{Sun Y, Chen F and Gou B C} 2011 {\em {J. Chem. Phys.}\/} {\bf \textbf{135}}
  124309

\bibitem{Cox1972}
{Daltabuit E and Cox D P} 1972 {\em {Astrophys. J.}\/} {\bf \textbf{177}} 855

\bibitem{Slater1930}
{Slater J C} 1930 {\em {Phys. Rev.}\/} {\bf \textbf{36}} 57

\bibitem{Slater1960}
{Slater J~C} 1960 {\em {Quantum Theory of Atomic Structure, Vol 1}\/} (New
  York, USA: McGraw-Hill)

\bibitem{Bethe1957}
{Bethe H A and Salpeter E E} 1957 {\em {Quantum Mechanics of One- and
  Two-Electron Systems}\/} (New York, USA: Academic Press)

\bibitem{Gatuzz2013}
{Gatuzz E {\it et al} } 2013 {\em {Astrophys. J}\/} {\bf \textbf{768}} 60

\bibitem{Lin2001}
{Lin S -H, Hsue C - S and Chung K T} 2001 {\em {Phys. Rev. A}\/} {\bf
  \textbf{64}} 012709

\bibitem{Lin2002}
{Lin S -H, Hsue C - S and Chung K T} 2002 {\em {Phys. Rev. A}\/} {\bf
  \textbf{65}} 032706

\bibitem{Ferland1998}
{Ferland G J, Korista K T, Verner D A, Ferguson J W, Kingdon J B and Verner E
  M} 1998 {\em {Pub. Astron. Soc. Pac.(PASP)}\/} {\bf \textbf{110}} 761

\bibitem{Ferland2003}
{Ferland G J} 2003 {\em {Ann. Rev. of Astron. \& Astrophys.}\/} {\bf
  \textbf{41}} 517

\bibitem{Kallman2001}
{Kallman T R and Bautista M A} 2001 {\em {Astrophys. J. Suppl. Ser.}\/} {\bf
  \textbf{134}} 139

\bibitem{Foster2012}
{Foster A R, Ji L, Smith R K and Brickhouse N S} 2012 {\em Astrophys. J\/} {\bf
  \textbf{756}} 128

\end{thebibliography}

\end{document}